\documentclass[aps,twocolumn,superscriptaddress,citeautoscript]{revtex4}

\usepackage{psfrag,graphicx}
\newcommand{\be}{\begin{equation}}
\newcommand{\ee}{\end{equation}}
\newcommand{\bra}[1]{\langle #1 |}
\newcommand{\ket}[1]{| #1 \rangle}

\newcommand{\ep}{\epsilon}

\begin{document}

\title{Simplified diagrammatic expansion for effective operator}

\author{Chang-Kui Duan}
\affiliation{Institute of Applied Physics and College of Electronic Engineering,
Chongqing University of Post and Telecommunication, Chongqing 400065, China}
\affiliation{Department of Physics and Astronomy, University of Canterbury, Christchurch,
  New Zealand}
\author{Yun-Gui Gong}
\author{Hui-Ning Dong}
\affiliation{Institute of Applied Physics and College of Electronic Engineering,
Chongqing University of Post and Telecommunication, Chongqing 400065, China}
\author{Michael F. Reid}
\affiliation{Department of Physics and Astronomy, University of Canterbury, Christchurch,
  New Zealand}
\date{\today}

\begin{abstract}
For a quantum many-body problem,
effective Hamiltonians that give exact
eigenvalues in reduced model space usually have different expressions, 
diagrams and evaluation rules from effective transition operators that give
exact transition matrix elements between effective eigenvectors in 
reduced model space. By modifying these diagrams slightly and
considering the linked diagrams for all the terms of the same order,
we find that the evaluation rules can be made the same for both
effective Hamiltonian and effective transition operator diagrams,
and in many cases it is possible to combine many
diagrams into one modified diagram. We give the rules to evaluate
these modified diagrams and show their validity.
\end{abstract}

\maketitle

\section{Introduction}
Effective Hamiltonian $H_{\rm eff}$ and transition operators $O_{\rm eff}$
 are commonly used in many {\it ab initio} many-body calculations\cite{HurF1993},
such as nuclear, atomic and molecular systems, and in phenomenological
calculations\cite{Lin1984,OleF1983,Bra1967,EllO1977,Bra1975,Bra1977,JorP1974,Fre1989,WanF1989a,WanF1991,WanF1989c,Kan1991,DuaR2002,AbeEE2003} of doped transitional metal ions, lanthanide and actinide
ions.  $H_{\rm eff}$ is defined to be an operator
acting on a restricted model space of handleable dimensions to give
upon diagonalization the exact eigenvalues and model space eigenvectors. 
For a time-independent transition operator $O$, $O_{\rm eff}$ may be introduced
that gives the same matrix elements between the model space eigenvectors of $H_{\rm eff}$
as the original operator $O$ between the corresponding true eigenvectors of $H$.
Calculations based on $H_{\rm eff}$ and $O_{\rm eff}$ have many advantages over
variational and other direct calculations based on $H$ and $O$, such as smaller bases,
less calculation effort, order by order approximations etc, and can also be used 
together with variational and other direct calculations\cite{Hur1993.} A recently review
can be found in \onlinecite{KilJ2003a}.

$H_{\rm eff}$ and $O_{\rm eff}$ are often constructed
by time-independent many-body perturbation theory (MBPT) with order-by-order
approximation, which can be represented with Goldstone diagrams, in analogy
to Feynman diagrams\cite{LinM1985}. Many expansions have been given, whose diagrammatic representations
usually contain not only connected but also  disconnected diagrams, which 
have the shortcoming of size inconsistency,  much more
computation efforts and tremendous number of diagrams. Nonetheless, There are 
common-used effective Hamiltonians, one hermitian and the other nonhermitian
are known to contain only connected diagrams.\cite{Hur1993} The rules to generate diagrams and
 to evaluate them are well-known. Factorization theorem is shown to be 
able to combine diagrams having the same set of vertexes and lines but different
relative orderings of vertexes, and hence reduces the number of high-order diagrams.\cite{Bra1967}
Compared to effective Hamiltonians, effective transition operators generally have
different algebraic forms, much more complicated diagrammatic representations, 
much greater number of diagrams  and different diagram evaluation rules. The 
hermitian (canonical) effective transition operator, which works together with
the hermitian (canonical) effective Hamiltonian, has been presented in detail
by Hurtubise and co-workers in a series of papers.\cite{HurF1993,Hur1993,Fre1989}
 Duan and Reid\cite{DuaR2001} constructed a
simple connected nonhermitian $O_{\rm eff}$ that works together with the connected
nonhermitian effective Hamiltonian and showed how to construct a connected expansion.
Since it is well-known that hermitian effective Hamiltonian up to third order can be
obtained from trivial symmetrization,\cite{Bra1975} and this can also be shown for effective transition operators
(up to third order in $V$, the perturbation in Hamiltonian), and also for higher-order calculations
usually coupled-cluster methods come into play, most researches required only limited order
diagram calculations of nonhermitian effective Hamiltonian and nonhermitian effective transition operators.
However, there are two problems with the effective operator methods: too many diagrams, and 
the rules to calculate energy denominators for effective transition operators
being different from those for effective Hamiltonian.\cite{Hur1993}

Our aim in this paper is to modify the diagram representation of many-body perturbation expansion of 
effective Hamiltonian and effective transition operators, so that
 the energy denominator rules for effective transition operators and effective
Hamiltonian are the same, and the number of diagrams is greatly reduced and becomes
handleable for third order terms of nonhermitian effective transition operators.
In section II we present our diagrammatic representation of many-body operators and the 
corresponding evaluation rules  for perturbation diagrams; in section III we 
illustrate the generalized factorization theorem for the effective transition operators;
and in Section IV we show how to denote  several diagrams of the same
set of vertexes and lines with only one diagram to group and reduce the number of diagrams.

\section{Modified diagrams and the evaluation rules}

We use the following algebraic representation for a general
 fermion n-body interaction $T_n(1,2,\cdots,n)$:
\begin{widetext}
\begin{eqnarray}
T_n 
    &=&    \frac{1}{n!^2}
  \sum\limits_{\alpha_1,\cdots,\alpha_n,\beta_1,\cdots, \beta_n}
    t_{\alpha_1\alpha_2\cdots\alpha_n,\beta_1\beta_2\cdots\beta_n} 
        a^+_{\alpha_1}\cdots a^+_{\alpha_n} a_{\beta_n}\cdots
        a_{\beta_1},\label{n_body_op}
\label{tn_eq}
\end{eqnarray}
where the coefficients
$t_{\alpha_1\alpha_2\cdots\alpha_n,\beta_1\beta_2\cdots\beta_n}$ 
are anti-symmetry under exchange of a pair of bra indexes or ket
indexes. It can be seen for one and two body interactions, the
coefficients are given as follows
\begin{eqnarray}
t_{\alpha,\beta} &=& \bra{\alpha(1)}T_1(1)\ket{\beta(1)},\\
\label{t1_coeff}
t_{\alpha_1\alpha_2,\beta_1\beta_2} &=&
\bra{\alpha_1(1)\alpha_2(2)}T_1(1,2)\ket{\beta_1(1)\beta_2(2)}-
\bra{\alpha_1(1)\alpha_2(2)}T_1(1,2)\ket{\beta_2(1)\beta_1(2)} ,
\label{t2_coeff}
\end{eqnarray}
\end{widetext}
where the indexes in bracket ``()'' means the indexes of the variables of 
the function. It is straightforward to give the coefficients for
$n$-body ($n\geq 3$)interactions. The diagrammatic representation of $T_n$ is:
\begin{equation}
\psfrag{Tn}{$T_n$}
\psfrag{A1}{$\alpha_1$}
\psfrag{A2}{$\alpha_2$}
\psfrag{An}{$\alpha_n$}
\psfrag{B1}{$\beta_1$}
\psfrag{B2}{$\beta_2$}
\psfrag{Bn}{$\beta_n$}
T_n=\mbox{\includegraphics[width=3cm]{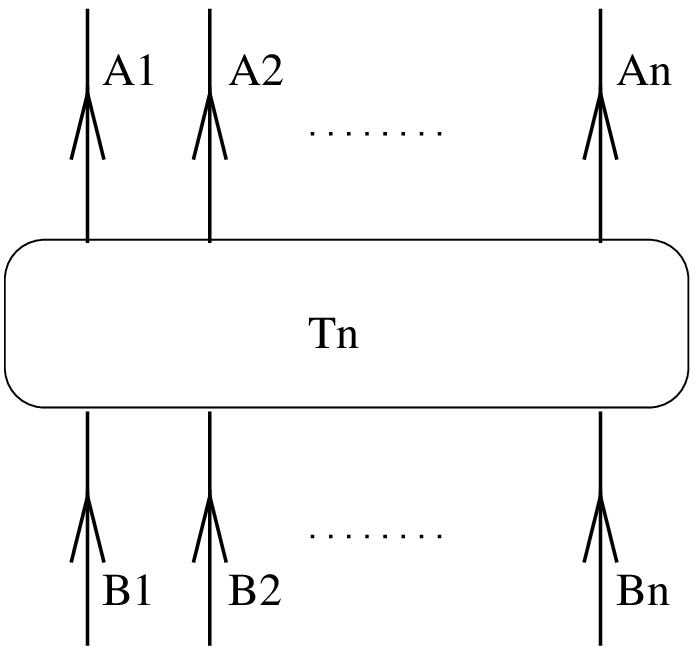}}
\label{tn_diagram}
\end{equation}
The diagram is invariant under $n!$ interchange of the $n$ out-going or
in-going lines. The number of symmetry operations is $(n!)^2$, which
contributes to the factor $1/(n!)^2$ as appeared in Eq.\ \ref{tn_eq}. For 
general close-shell reference ``vacuum'' state where there are both
particle and hole lines, the number of inequivalent diagrams for $T_n$ 
are $(n+1)^2$, corresponding to the $n+1$ possibilities for the
in-going lines (0, 1, $\cdots$, $n$ holes) and out-going
lines. Particle line and hole line are inequivalent, and hence the symmetry
factor changes accordingly for each diagram.

\begin{figure}
\psfrag{Veff}{\small $V_{\rm eff}$}
\psfrag{Vb}{\small $V_1$}
\psfrag{Vc}{\small $V_2$}
\psfrag{Va}{\small $V_3$}
\psfrag{3va1}{(a1)}
\psfrag{3va2}{(a2)}
\psfrag{3vb}{(b)}

\mbox{\includegraphics[width=4.5cm]{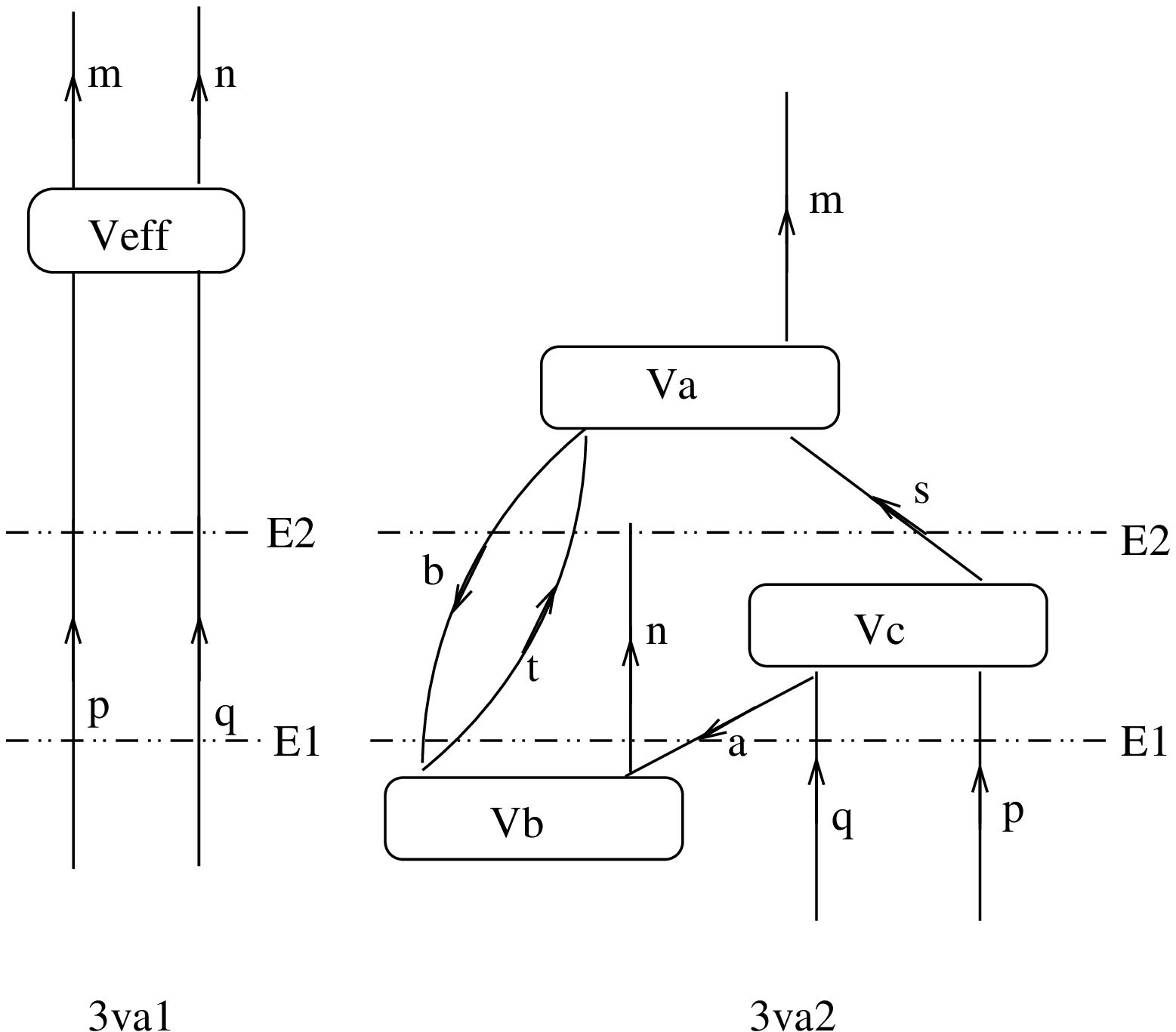}}
~~~~~~
\mbox{\includegraphics[width=3cm]{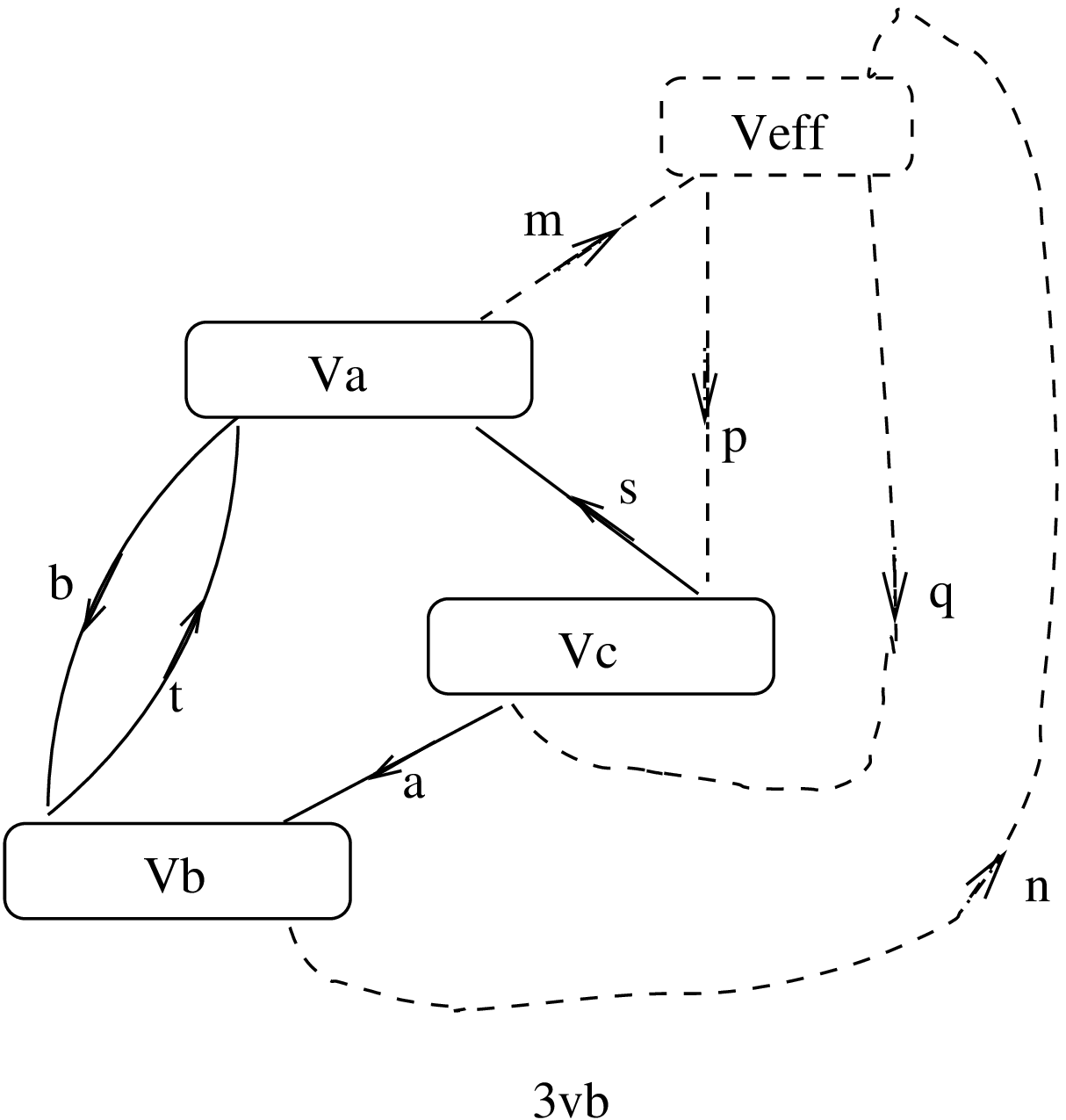}}
\caption{\label{vvv}
A two-body third order diagram of non
hermitian Hamiltonian, where (a2) is the original diagrams which
contributes to the effective interaction showed by diagram
(a1), (b) is our modified representation for this diagram,
where out-/in- going lines are changed into dashed line and 
end at dashed interaction vertex $V_{\rm eff}$.
}
\end{figure}

\begin{figure}
\psfrag{2vd1}{(a1)}
\psfrag{2vd2}{(a2)}
\psfrag{2vdb}{(b)}
\mbox{\includegraphics[width=5cm]{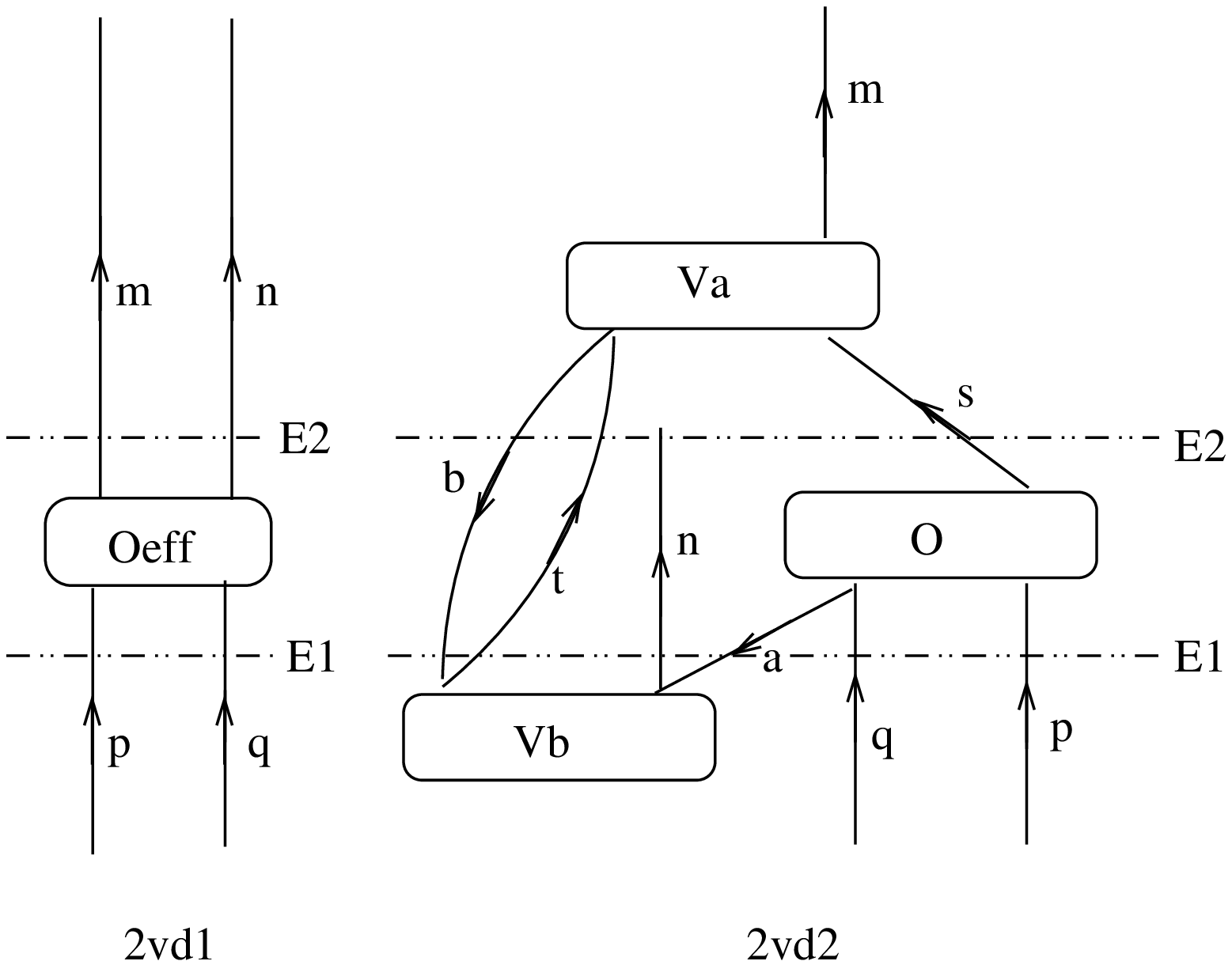}}
~
\mbox{\includegraphics[width=3cm]{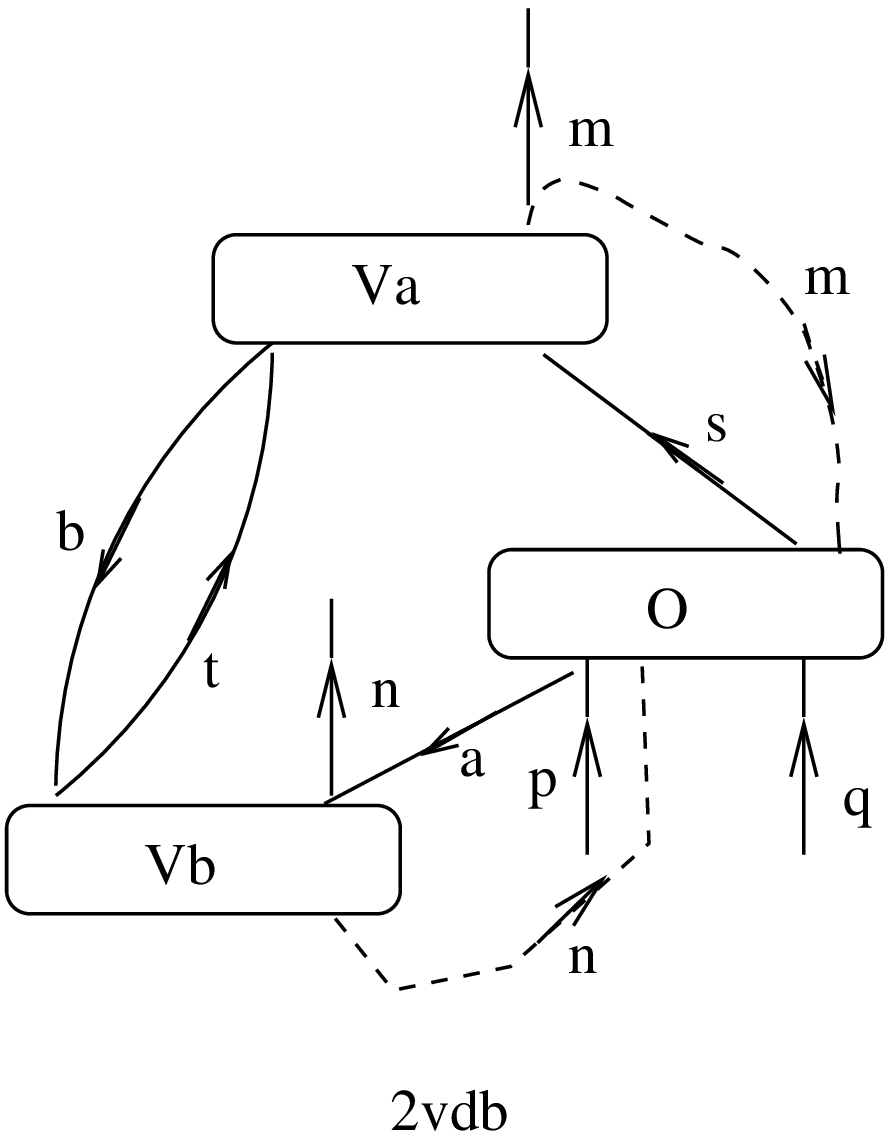}}
\caption{\label{vdv}
A second order (in $V$) diagram of effective transition
operator $O_{\rm eff}$, where $V_a$ and $V_b$ are interactions contained
in original Hamiltonian and $O$ is the original transition operator. 
}
\end{figure}

We need to modify the diagrams slightly so that the evaluation of energy
denominators for effective operator is the same as that for
non-hermitian Hamiltonian and is simple in general. This simplification helps us using
generalized factorization theorem and simplifies diagrams developed in
the next two sections. The modifications are shown in Fig.\ref{vvv}
and Fig.\ref{vdv}.

The general rules for evaluating the diagrams of non hermitian Hamiltonian
 have been
summarized by Brandow\cite{Bra1967} and Lindgren and Morrison\cite{LinM1985}. The rules
of evaluating the diagrams for second order perturbative expansion of
 effective operator have been given by Hurtubise and Freed
\cite{HurF1994}. With the new notation (\ref{tn_eq})-(\ref{tn_diagram}),
 the revised more general diagram evaluation rules can be summarized as follows:
 1) draw only one diagram\cite{Bra1967} for any one $T_n$ operator;
 2) express the evaluation result in the form 
 Eq. (\ref{tn_eq}) and the evaluation becomes calculation 
 of $t_{\alpha_1,\alpha_2,\cdots;\beta_1,\beta_2,\cdots}$ coefficients;
 3) in coefficient $t_{\alpha_1,\cdots,\alpha_m;\beta_1,\cdots,\beta_m}$, includes
 a factor $(m!) ^2/ (i_1!i_2!\cdots)$, where $m$ ingoing lines and $m$ outgoing lines
 are participated into sets with $i_1$, $i_2$, $\cdots$, equivalent lines, with $i$ lines
 being equivalent if and only if they all start at the same 
 interaction (or no beginning interaction), and point to the same 
 ending interaction (or no ending interaction);
 4) include an over-all sign factor of $(-1)^{l+h}$,\cite{Bra1967} where $l$ is the number of 
 closed loops and $h$ the number of downgoing or "hole" line segments;
 5) For each vertex, multiple the antisymetrized matrix elements as given in Eq. 
 (\ref{t1_coeff}) and (\ref{t2_coeff}).
 6) for a n-vertex diagram (suppose the vertex are labeled increasing
  from the bottom as 1, 2, $\cdots$, $n$), include the products of energy denominators calculated
  just above the vertex 1, 2, $\cdots$, $n$, where energy denominators are 
  calculated by energy sum of all down going line taking away the energy sum of all
  upgoing lines, as show in Fig. \ref{vvv};
  7) sum each upgoing line independently over all particle states, 
  and each downgoing line independently over all hole states.
  The exclusion-violating terms which arise from these independent summations
  must be included.\cite{Bra1967}

 The energy denominator can be calculated for diagram Fig.1 (a2) as
\begin{widetext}
\begin{eqnarray}
&&(E_2(1)-E_2(2))(E_1(1)-E_1(2))\nonumber\\
 &=& [(\ep_p+\ep_q)-(-\ep_a-\ep_b+\ep_t+\ep_n
+\ep_p+\ep_q)]
[(\ep_p+\ep_q)-(-\ep_b+\ep_t+\ep_s+\ep_n)] \\
&=& (\ep_a +\ep_b-\ep_t-\ep_n)(\ep_b+\ep_p+\ep_q-\ep_t-\ep_n-\ep_s),
\end{eqnarray}
where $E_i(j)$ is the energy for diagram Fig.\ \ref{vvv}(ai) calculated
at the indicated line by the energies of up-going lines
taking away the energies of down-going lines. Diagram Fig.\ \ref{vvv} (a2) can
be written as
\begin{eqnarray}
    {\rm Fig}.\ \ref{vvv} (a2) 
&=& \frac{1}{2!^2}\sum\limits_{mnpq} a^+_ma^+_na_qa_p 
   \left (-(2!)^2/2!\sum\limits_{abts}
 \frac{ (V_c)_{tn,ba}(V_b)_{sa,pq}(V_a)_{mb,st}}
  { (\ep_a +\ep_b-\ep_t-\ep_n)
    (\ep_b+\ep_p+\ep_q-\ep_t-\ep_n-\ep_s)
  }
  \right ),
\end{eqnarray}
\end{widetext}
where the minus sign comes from the numbers of core solid lines,2, 
and close cycle,1, adding up to odd number 3. the factor $(2!)^2/2!$ 
is due to the fact that the effective operator
is two-body with equivalent lines $p$ and $q$.
 
The calculation of energy denominator can be put in an more systematic way
as shown in Fig. \ref{vvv}b:  the energy denominator for $i^{\rm th}$ vertex,
is just the negative of net outflow energy ($E_{\rm noe}$) of a loop
 enclosing vertex 1, 2, $\cdots$, $i$, i.e., 
\begin{equation}
-E_{\rm noe} (V_1V_2\cdots V_i) = - (E_{\rm noe} (V_1) + E_{\rm noe} (V_2) +
\cdots + E_{\rm noe}(V_i)).
\end{equation}
Therefore, the energy denominator factor for the $n-$ vertex diagram can be
written as
\begin{widetext}
\begin{eqnarray}
D&=& (-1)^{n-1} E_{\rm noe}(V_1)E_{\rm noe}(V_1V_2) \cdots E_{\rm noe}(V_1V_2\cdots V_{n-1})\\
 &=& (-1)^{n-1} E_{\rm noe}(V_1)[  E_{\rm noe}(V_1)+  E_{\rm noe}(V_2)] \cdots
    [  E_{\rm noe}(V_1) +   E_{\rm noe}(V_2) + \cdots   E_{\rm noe}(V_{n-1})].
\label{Vn_denom}
\end{eqnarray}
\end{widetext}

For Hermitian effective transition operators and Hamiltonians, the rules to
evaluate diagrams have been given by Hurturbise and
Freed\cite{HurF1994}, which is different from Lindgren and
Morrison's\cite{LinM1985} in energy denominators and contains
algebraic factors. When we draw 
vertex of $O_{\rm eff}$ at the same level as the vertex of $O$ in
perturbation diagrams, as shown in Fig.\ \ref{vdv}(a), the 
rules of calculating denominators become exactly the same as those for
the non-hermitian Hamiltonian. We modify Fig.\ \ref{vdv}(a) accordingly to
give Fig.\ \ref{vdv}(b). In Fig.\ \ref{vdv}(b) we save
the free lines for the sake of calculating matrix elements. Note that the
free lines should not be included in the calculation of denominators and 
the dashed lines should not be included in the calculation of matrix
elements.  Diagram Fig.\ \ref{vdv}(a)
can be calculated as
\begin{widetext}
\begin{equation}
  {\rm Fig}.\ \ref{vdv}(b) 
= \frac{1}{4}\sum\limits_{mnpq}a^+_ma^+_na_qa_p
\left (-2 \sum\limits_{abst}
 \frac{(V_3)_{mb,st}(V_1)_{nt,ab}O_{as,pq}}
      {(\ep_a+\ep_b-\ep_t-\ep_n)(\ep_b+\ep_m-\ep_t-\ep_s)}
\right )
\end{equation}
\end{widetext}
It can been seen that the rules to calculate the denominators are the same for
 the two types of diagrams, which are given 
in Eq.\ \ref{Vn_denom}.

\section{Generalized factorization theorem for the modified diagrams}

When several diagrams have the same set of vertexes and the same line
directions and line types but different relative 
orderings of vertexes, they have the same matrix elements, factors and
signs but different denominators. If all possible relative
orderings are included, the results can be expressed by means of a
single diagram with the denominators determined independently for each
part of the diagrams\cite{Bra1967}. This is the factorization theorem.
Lindgren gave a illustration and proof\cite{Lin1974}. However it was
not trivial to show that this theorem also holds for Hurtubise  and
Freed's rules of evaluation of  denominators for  effective interaction
diagrams. Here, the rules to evaluate denominators for our modified diagrams
are the same for both effective Hamiltonian and effective operator,
showing trivially that factorization theorem also holds for effective
operator.

\begin{figure}
\includegraphics[height=4cm,width=4cm]{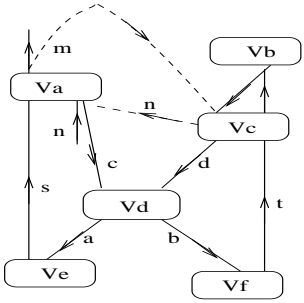}
\caption{\label{GFT}
An example connected diagram with disconnected part at the bottom and 
on the top to illustrate factorisation theorem.
}
\end{figure}

 A diagram to illustrate this theorem is given in Fig.\ \ref{GFT}, where 
 $V_e$ and $V_f$ are two disconnected parts at the bottom, and $V_a$ and 
 $V_bV_c$ are two disconnected parts on the top. If one
 lowers $V_e$ relative to $V_f$, the vertexes and all the lines
 (direction and types) do not change, but the denominator from the
 lowest two interactions change from $E_{\rm noe}(V_f) [E_{\rm noe}(V_e) +
 E_ {\rm noe}(V_f)]$ to $E_ {\rm noe}(V_e) [E_ {\rm noe}(V_f) + E_ {\rm noe}(V_e)]$. By
 adding those two diagrams together, one gets a denominator 
$ E_{\rm noe}(V_f) E_{\rm noe}(V_e)$, which is the same as calculating the lowest
 two disconnected interaction separately.  The denominator for $V_b$ is
 $-E_{\rm noe}(V_a V_c  V_dV_eV_f) = E_{\rm noe} (V_b)$, which follows from the
 fact that all  $E_{\rm noe}$'s for all vertexes add up to
 zero. Similarly, the  denominator from the highest three interactions
 is $ E_{\rm noe}(V_b) E_{\rm noe} (V_bV_a) E_{\rm noe} (V_b V_a
 V_c)$ for the ordering showed in Fig.\ \ref{GFT}. There are two other
 diagrams with different orderings of  $V_a$ relative to $V_bV_c$. By
 adding all the three diagrams up, one gets, similar to the case of the
 lowest two interactions, a denominator from the highest three
 interactions  $ E_{\rm noe}(V_a) E_{\rm noe}(V_b) [
 E_{\rm noe}(V_bV_c)]$. The total denominator calculated this way for the sum of those 
 diagrams with all the possible orderings is $ E_{\rm noe}(V_a)
 E_{\rm noe}(V_b) E_{\rm noe}(V_b V_c)E_{\rm noe}(V_e) E_{\rm noe}(V_f)$. If one
 uses factorization theorem for both the upper disconnected part and the 
 lower disconnected part, one gets exactly the same denominator much
 more straightforward. For completeness, we give the result for this
 diagram as follows:
\begin{widetext}
\begin{eqnarray}
Fig.\ \ref{GFT} &=& \sum\limits_{mn}a^+_ma_n \left ( (-1)^6
     \sum\limits_{abcde,stu} \frac{(V_a)_{mc,sn} (V_b)_{e,u}
     (V_c)_{ud,te}(V_d)_{ab,cd}(V_e)_{s,a}(V_f)_{tb}}
    {E_{\rm noe}(V_a) E_{\rm noe}(V_b) E_{\rm noe}(V_b V_c)E_{\rm noe}(V_e)
      E_{\rm noe}(V_f) } \right ) 
\label{gfta}\\
&=& \sum\limits_{mn}a^+_ma_n \left ( (-1)^8
     \sum\limits_{abcde,stu} \frac{(V_a)_{mc,ns} (V_b)_{e,u}
     (V_c)_{ud,et}(V_d)_{ab,cd}(V_e)_{s,a}(V_f)_{tb}}
   {E_{\rm noe}(V_a)
 E_{\rm noe}(V_b) E_{\rm noe}(V_b V_c)E_{\rm noe}(V_e) E_{\rm noe}(V_f) } \right ).
\label{gftb}\\
&=& \sum\limits_{mn}a^+_ma_n \left (
     \sum\limits_{abcde,stu} \frac{(V_a)_{mc,ns} (V_b)_{e,u}
     (V_c)_{ud,et}(V_d)_{ab,cd}(V_e)_{s,a}(V_f)_{tb}}
   {(\ep_c+\ep_m-\ep_n-\ep_s)(\ep_e-\ep_u)(\ep_n+\ep_d-\ep_t-\ep_m) 
    (\ep_e-\ep_s)(\ep_b-\ep_t) }\right ).
\end{eqnarray}
\end{widetext}
The sign $(-1)^6$ comes from the fact that there are five backwards
solid internal lines and one loop ($edbtue$). If one switches the
entering points of $s$ and
$n$ at vertex $V_a$, and $e$ and $t$ at vertex $V_c$, then one gets the
equivalent diagram with three loops. The result for this diagram is
given in (\ref{gftb}). 

\section{Diagrams with the same set of vertexes and line directions and
  notations} 

\begin{figure}
\psfrag{equals}{$=$}
\includegraphics[width=8.9cm]{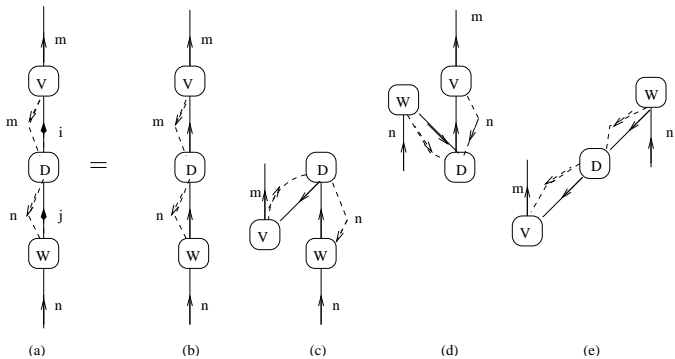}
\caption{\label{topo}
An example to demonstrate the combination of several diagrams with the same
set of vertexes, line connections and directions into one diagram. Here
(b) has two particle-state internal lines, (c) and (d) each has one
 hole-state internal line and one particle-state internal line, (e) 
 has two hole-state internal lines,
 and (a) is a diagram used to denote all the four diagrams (b) to (e).
}
\end{figure}

The diagrams in Fig.\ \ref{topo}. (b), (c), (d) and (e)
have the same set of vertexes, line connections and directions. 
One can simply denote the four diagrams as diagram (a), where $i$
and $j$ can be both core
lines and virtual lines. Hence the matrix elements, the symmetry factors
and the $E_{\rm noe}$'s have the same form. The denominators are just
different combinations of $E_{\rm noe}$'s. One cannot combine the
denominators together the same way as in the factorisation theorem since
the orbital types for different diagrams are different. However, this
does not prevent us from using (a) to represent all the four diagrams
as long as one is cautious about the relations between the relative
orderings of vertexes and the internal line types in evaluating diagram
(a).
\begin{eqnarray}
{\rm Fig.\ \ref{topo}} &=& \sum\limits_{mn}a^+_ma_n \left (
   \sum\limits_{i,j} \eta_1(i,j)\frac{V_{mi}D_{ij}V_{jn}}{\eta_2(i,j)
     E_{\rm noe}(V) E_{\rm noe}(W)} \right )\nonumber\\
 &=& \sum\limits_{mn}a^+_ma_n \left (
   \sum\limits_{i,j}\frac{V_{mi}D_{ij}V_{jn}}{
    (\ep_n-\ep_j)(\ep_m-\ep_i)}
 \right ),
\end{eqnarray}
where $\eta_1(i,j)=1,-1,-1,1$ and $\eta_2(i,j)= -1,1,1,-1$ are the sign
contributed by core orbitals and denominators respectively. Note that
factorisation theorem has been applied in the evaluation of diagrams (c) and
(d). 

Denote the line entering $V_1$ and then passing sequentially
$V_2,\cdots,V_{n-1}$ and finally going out from $V_n$ as
$(V_nV_{n-1}\cdots V_2 V_1)$, the loop passing sequential
$V_1,V_2,\cdots,V_n, V_1$ and pointing from $V_1$ to $V_2$ as
$(V_1V_n\cdots V_2 V_1)$, and the ordering that $A$ is higher then $B$
as $[AB]$. It can be 
seen that diagram Fig.\ \ref{topo}a is uniquely specified by
$(VDW)$ and diagram Fig.\ \ref{topo}b-e are
$(VDW)[VDW]$,$(VDW)[DVW+DWV]$,$(VDW)[VWD+WVD]$ and
$(VDW)[WDV]$. As has been argued above, one only needs to draw one
diagram to work out the values of all those similar diagrams by using
the denominator rule Eq.\ \ref{Vn_denom} and other standard rules for
matrix elements and factors. The application of these new notations to
one and two-photon transitions can greatly reduce the number of diagrams.
This will be presented elsewhere latter.

\section{conclusion}

We have modified the diagrams for
non-hermitian effective Hamiltonians and effective operators so that
the rules to calculate the  denominators are the same for the two types
of diagrams. Expressed with these modified diagrams, proof of factorization
theorem for effective transition operators becomes trivial. By denoting the
class of diagrams of the same set of vertexes, line connections and directions
with one diagram, the number of diagrams can be greatly reduced and hence making
the high order contributions to effective operators more handleable.

\bibliography{306434JCP}
\end{document}